\documentclass[10pt, twocolumn]{article}

\usepackage{amsmath}
\usepackage{appendix}
\usepackage{booktabs}
\usepackage{fancyhdr}
\usepackage{fullpage}
\usepackage{graphicx}
\usepackage{hyperref}
\usepackage{stfloats}		
\usepackage[footnotesize]{caption}

\begin{document}

\title{Perturbed and Permuted \\ \large Signal Integration in Network-Structured Dynamic Systems}
\author{Dennis Wylie}
\maketitle

\thispagestyle{empty}
\renewcommand{\headrulewidth}{0.4pt}
\pagestyle{fancy}

\begin{abstract}
Biological systems (among others) may respond to a large variety of distinct external stimuli, or signals. These perturbations will generally be presented to the system not singly, but in various combinations, so that a proper understanding of the system response requires assessment of the degree to which the effects of one signal modulate the effects of another. This paper develops a pair of structural metrics for sparse differential equation models of complex dynamic systems and demonstrates that said metrics correlate with proxies of the susceptibility of one signal-response to be altered in the context of a second signal. One of these metrics may be interpreted as a normalized arc density in the neighborhood of certain influential nodes; this metric appears to correlate with increased independence of signal response.
\end{abstract}

\section{Introduction}
Biological signaling pathways frequently intersect one another, leading to the phenomenon of cross-talk, wherein the effect of one signaling pathway influences the activity of another. For example, during Drosophila development, the epidermal growth factor receptor (EGFR) and Notch signaling pathways interact both within and between cells, producing highly context-specific responses to their respective signals during the formation of spatially structured organs including the eye \cite{doroquez2006signal}. The interactions between these pathways range from antagonistic to cooperative across different processes and at different times within a given process, resulting in an intricate integration of signal response.

The concept of signal interaction embodied in this example can be generalized to less obvious biological examples as well: for instance, it has been suggested that cell-to-cell variation between embryonic stem (ES) cells may result in differential responses to differentiation signals, thereby permitting some ES cells to undergo lineage specification while others remain pluripotent \cite{silva2008capturing}. Regarding the differences in molecular population and configuration constituting such ES cellular variation as random perturbations (or more abstractly, signals) to the biochemical state, this phenomenon may also be described as the modulation by one signal of the systemic response to another.

A key question in such cases is to what extent the presence and magnitude of one signal interferes with or reinforces the effects of the other. Perhaps the simplest possibility is linear superposition, in which the system essentially responds with the sum of the responses it would have to the two signals if presented separately --- i.e., despite making use of common signaling components, the two signals act in a fundamentally independent manner. Such ``signal independence'' is, however, inconsistent with many properties ascribed to the interlinked networks of biological signaling pathways. For instance, while an AND gate of sorts could be constructed with linear superposition of two signals if the effect of either one alone was below threshold, in the presence of noisy signals of widely varying magnitudes (more the rule than the exception in biological signaling), a particularly high magnitude single signal would lead to activation by itself. Meanwhile, even the complexity of a simple XOR (exclusive-or) gate would be impossible with linear signal superposition/signal independence.

Previous work by Wylie \cite{wylie2009linked} suggests that the network topology of dynamic systems plays a key role in determining the integration of multistable dynamic ``switches:'' specifically, sparse networks of relatively homogenous node degree were found to be more favorable to switch integration than were dense or scale-free networks. Here we investigate whether similar ideas might be applied to wider class of signal integration phenomena not necessarily involving multistability.

\section{Terminology and Notation}
\label{sec:terminology}

\begin{figure*}[!t]
\centering
\fbox{\begin{minipage}{\textwidth}
	\centering
	\includegraphics[width=0.90\textwidth]{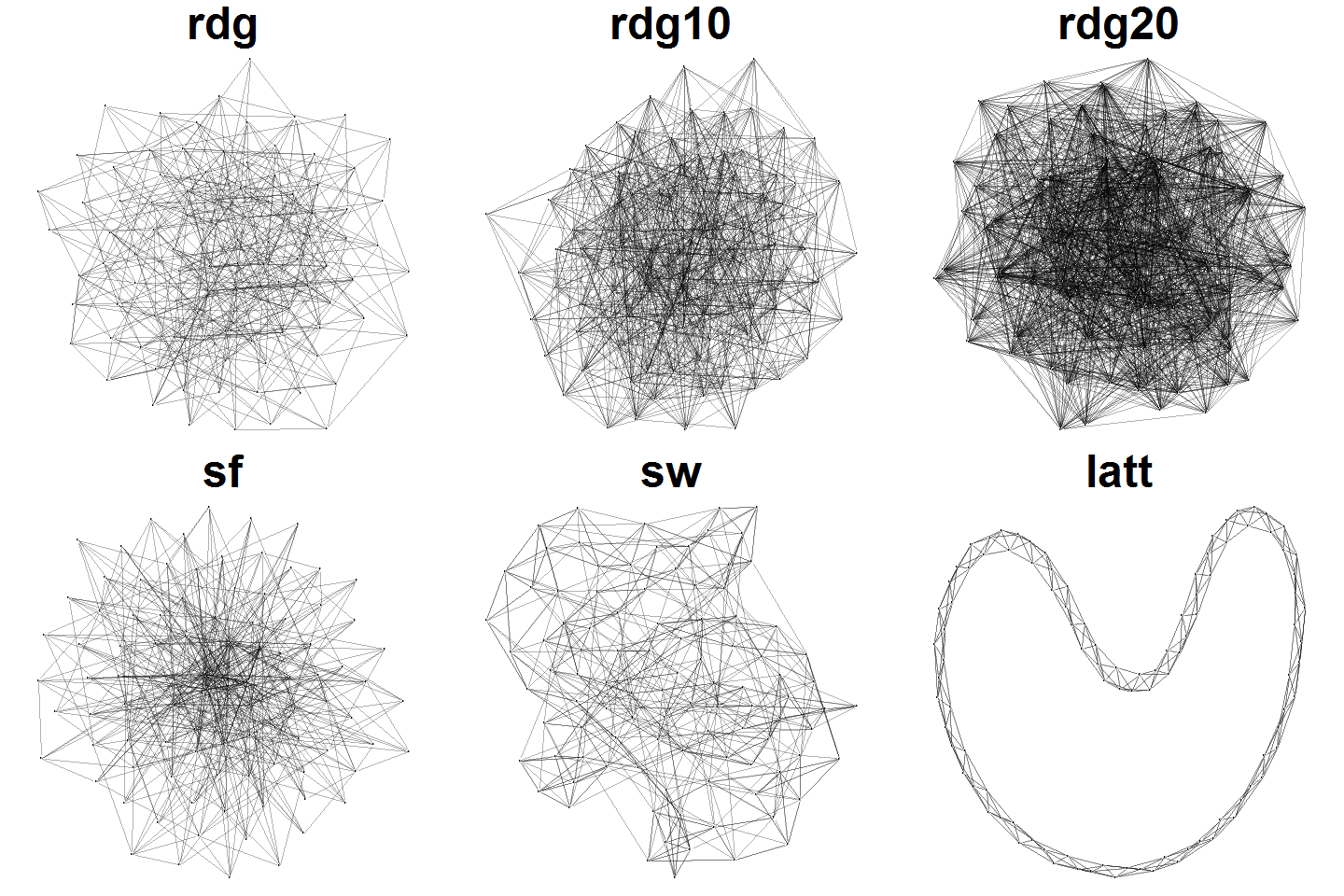}
	\caption{Characteristic network structures: \textbf{rdg} stands for random digraph, \textbf{sf} indicates scale-free, \textbf{sw} and \textbf{latt} indicate small-world digraphs with rewiring probabilities \(p_\text{rw}=0.1\) and \(p_\text{rw}=0\), respectively. Numerical postfixes indicate average node in- and out-degree; where not indicated, average degree is 6. Arc directionality suppressed for visual clarity.}
	\label{fig:net plots}
\end{minipage}}
\end{figure*}

We consider deterministic nonlinear dynamic systems with steady state at the origin, so that
\begin{align}
	\label{eq:general system}
  \frac{dx_i}{dt}
  &= f_i (\mathbf{x}) \\
  &= \sum_j {A_{ij} x_j}
  + \sum_{j,k} {B_{ijk} x_j x_k}
  + O( x^3 ) \nonumber
\end{align}
For the purposes of this paper, we will restrict ourselves to quadratic systems for which all third-order and higher terms vanish, so that the linearization matrix \(A\) and the three-index array \(B\) (which, without loss of generality, we assume is symmetric with respect to its 2\textsuperscript{nd} and 3\textsuperscript{rd} indices) totally determine the dynamics. We will also write equation \eqref{eq:general system} as
\begin{equation}
	\label{eq:simplified general system}
	\frac{d\mathbf{x}}{dt}
	= A\mathbf{x}
	+ B(\mathbf{x}, \mathbf{x})
\end{equation}

Models of large biological systems are generally sparse in the sense that most entries \(A_{ij}\) in the matrix \(A\) vanish. We can thus associate a network structure (more precisely, a directed graph in which ``one-loop'' arcs from a node \(i\) to itself are allowed) \(G = (V_G , E_G)\) with the system by
\begin{equation}
	(i \rightarrow j) \in E_G
	\text{ iff } A_{ji} \neq 0
	\text{ (Note index order) }
\end{equation}
We will assume here that the quadratic array \(B\) is consistent with the network structure \(G\), in the sense that
\begin{equation}
	B_{ijk} = 0	\text{ unless }
	(j \rightarrow i), (k \rightarrow i) \in E_G
\end{equation}
which is necessary if the linearized system structure \(G\) is to be stable to small perturbations of the form
\begin{equation}
	\mathbf{f}(\mathbf{x})
	\mapsto
	\mathbf{f}(\mathbf{x}) + \Delta \mathbf{c}
\end{equation}
as is discussed further in section \ref{sec:uvar}. The network structures \(G\) considered in this work are generally random digraphs, scale-free digraphs characterized by high variance of node (in- and out-) degree, and small-world digraphs (including lattice digraphs) characterized by high clustering. (Here in-/out-degree are defined ignoring both one-loop arcs and arc weights.) Figure \ref{fig:net plots} offers visualizations of some characteristic structures.

\section{Topological Properties of Characteristic Polynomial}

\begin{figure}[!t]
\fbox{\begin{minipage}{0.48\textwidth}
	\centering
	\includegraphics[width=\textwidth]{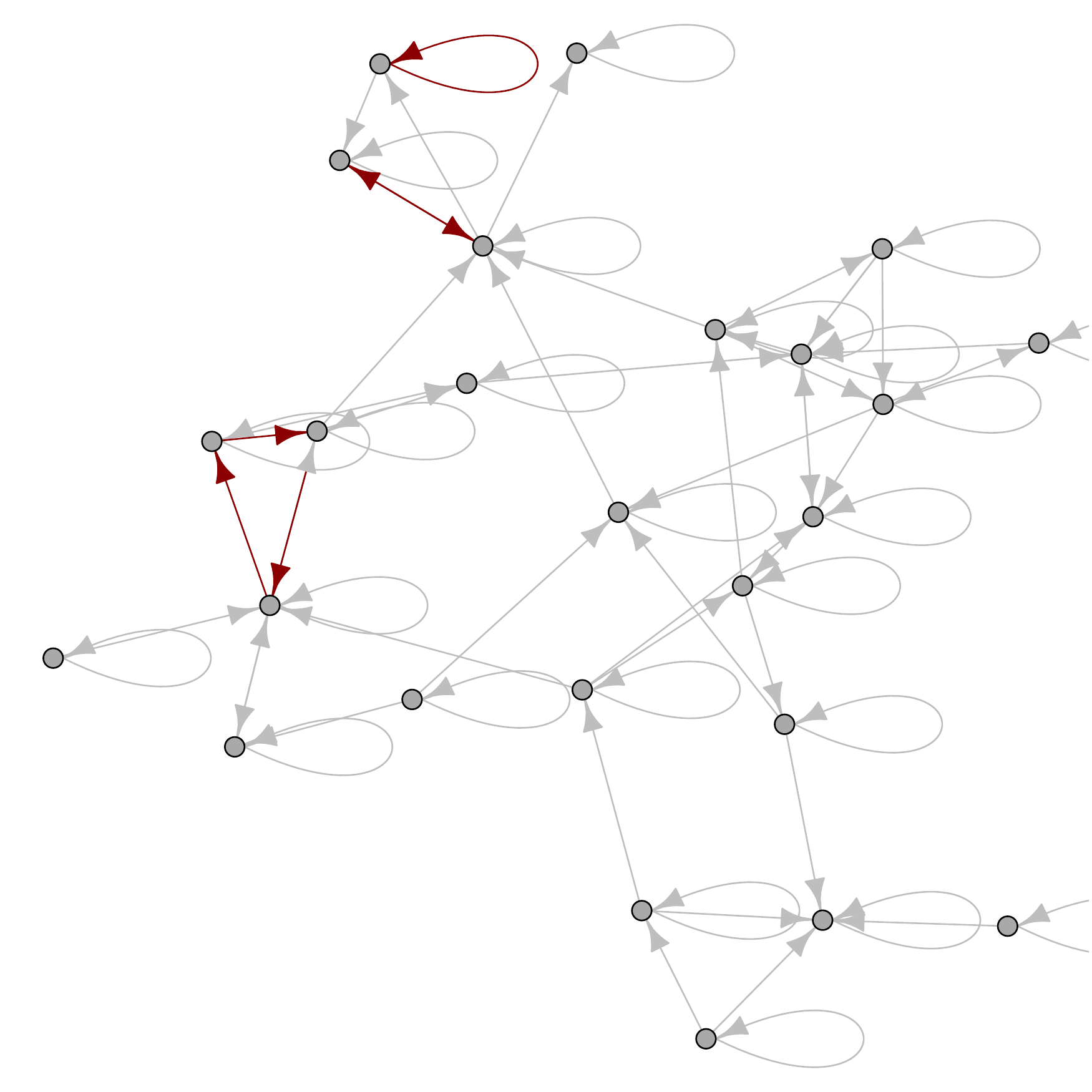}
	\caption{Arc set representation of permutation \(K\)=(123)(45)(6), with \(k=6\) and \(c(K)=3\), is highlighted in dark red.}
	\label{fig:permutation}
\end{minipage}}
\end{figure}

We here refer to the characteristic polynomial associated with the linearization matrix \(A\),
\begin{equation}
	\label{eq:char poly}
	\text{det}(A - \lambda I)
	= (-1)^{n-1}
	\sum_{k=0}^n {F_{n-k} \lambda^k}
\end{equation}
as the characteristic polynomial of the system described by equation \eqref{eq:general system}. The notation \(F_k\) used in equation \eqref{eq:char poly} is intended to evoke the interpretation of the coefficients of the characteristic polynomial as the ``feedback at length \(k\)'' in the (weighted) network \(G\). This interpretation stems from the relationship \cite{schneider1989matrices, puccia1985qualitative}
\begin{equation}
	\label{eq:feedback notation}
	F_k = \sum_{K \in \Theta_k} {\left[
		(-1)^{c(K)+1}
		\prod_{(i \rightarrow j) \in K} {A_{ji}}
	\right]}
	= \sum_{K \in \Theta_k} {w_K}
\end{equation}
where \(\Theta_k\) is the set of all permutations of \(k\) integers chosen from the set \(\{1,2,\ldots,n\}\). Here we regard any particular permutation \(K\) chosen from \(\Theta_k\) as a set of arcs \((i \rightarrow j)\) corresponding to the mappings of individual elements by the permutation operation (illustrated in figure \ref{fig:permutation}). Considering the standard cycle representation of permutation groups \cite{landin1989introduction}, it is apparent that, considered graphically, \(K\) will generally consist of a definite number \(c(K)\) of node-disjoint cycles, the sum of whose lengths is \(k\). We define here the weight \(w_K\) of the permutation \(K\) with respect to the system \(A\) to be the product of the arc weights \(A_{ji}\) associated with the arcs \((i \rightarrow j)\) making up \(K\) (times the sign factor \((-1)^{c(K)+1}\)), so that \(F_k\) is simply the sum of the length-\(k\) permutation cycle weights. This is the basis for considering \(F_k\) as a measure of feedback (of length \(k\)) in the linearized system \(A\).

\section{Variation under System Perturbation}
\label{sec:uvar}
We take signal inputs to our dynamic systems in the form of perturbations to the system dynamics
\begin{equation}
	\label{eq:pert signal}
	\mathbf{f}(\mathbf{x})
	\mapsto \mathbf{f}(\mathbf{x}) + \Delta \mathbf{c}
\end{equation}
In the context of a biochemical model, signals described by equation \eqref{eq:pert signal} might consist of the steady input and/or removal of a given set of chemical species.

For small perturbations, the root \(\Delta \mathbf{y}\) (\(=\Delta \mathbf{y}^{(1)} + \Delta \mathbf{y}^{(2)} + \text{ higher order terms}\)) of the function (\(\mathbf{f} + \Delta \mathbf{c}\)) is given to first order in \(\Delta \mathbf{c}\) by
\begin{equation}
	\label{eq:first order dy}
	\Delta \mathbf{y}^{(1)}
	= -A^{-1} \Delta \mathbf{c}
\end{equation}
If the system is perturbed by two distinct perturbations \(\Delta \mathbf{c}\) and \(\delta \mathbf{c}\) --- representing here two incoming signals --- the first order root shift response (\(\Delta \mathbf{y}^{(1)} + \delta \mathbf{y}^{(1)}\)) will obviously be a linear superposition of the responses to the two individual signals input singly. We then consider also the second order mixed terms,
\begin{equation}
	\label{eq:second order dy}
	\Delta \delta \mathbf{y}^{(2)}
	= -2 A^{-1} B(\Delta \mathbf{y}^{(1)}, \delta \mathbf{y}^{(1)})
\end{equation}
As the only terms we will be interested in of greater than first order will be mixed terms of the \(\Delta \delta\) form, we will henceforth drop the parenthetical superscripts.

We here consider two distinct outputs, in the form of changes to the steady-state properties of the dynamic system under consideration, resulting from the signal inputs \(\Delta \mathbf{c}\) and \(\delta \mathbf{c}\). These are the coefficients of the characteristic polynomial \(F_k\) (for brevity, we will generally refer simply to ``the characteristic polynomial \(F_k\)'') and the eigenvalues \(\lambda\) (particularly the ``least stable'' eigenvalue(s) \(\lambda_{\text{ls}}\) of largest real part \(\text{Re}(\lambda)\)). These particular outputs were chosen for their close relationship with system stability, as well as for some degree of analytic convenience.

One simple metric for quantifying the degree to which a signal \(\Delta \mathbf{c}\) applied to the system \(\mathbf{f}\) may be expected to change the effects of a randomly distributed second signal \(\delta \mathbf{c}\) with regard to a particular output function \(U\) is the Pearson correlation
\begin{align}
	\label{eq:uvar correlation}
	& \text{Corr}_{\delta} \left( \delta U + \Delta \delta U, \delta U \right)
	\\ & =  1
	+ \frac{\left\langle \left\langle \delta U \, \Delta \delta U \right\rangle \right\rangle_{\delta}^2}{2 \left\langle \left\langle [\delta U]^2 \right\rangle \right\rangle_{\delta}^2}
	- \frac{\left\langle \left\langle [\Delta \delta U]^2 \right\rangle \right\rangle_{\delta}}{2 \left\langle \left\langle [\delta U]^2 \right\rangle \right\rangle_{\delta}}
	+ \text{H.O.T.} \nonumber
\end{align}
where the correlation and (co)variances are all taken with respect to the distribution of \(\delta \mathbf{c}\) (the notation \(\left\langle \left\langle XY \right\rangle \right\rangle\) indicates the covariance of the random variables X and Y). Equation \eqref{eq:uvar correlation} is a simple lowest order expansion of the Pearson correlation in \(\Delta \delta U\). It is worth noting that since we are considering a quantity which clearly depends on the overall scale of the perturbations \(\delta U\), \(\Delta U\), and \(\Delta \delta U\), and since we would like to compare this quantity across systems with very different structures, we must consider how to properly normalize the signals \(\delta \mathbf{c}\) and \(\Delta \mathbf{c}\). For the purposes of this work, signals were normalized by the root-mean-square average magnitudes of the first-order shifts \(\delta \lambda_{\text{ls}}\) to the least-stable (largest real-part) eigenvalues of the linearization matrix \(A\) (see sections \ref{sec:numerical arc analysis} - \ref{sec:numerical analysis pert}).

To apply equation \eqref{eq:uvar correlation} to our cases of interest (\(U=F_k\) or \(U=\lambda_{\text{ls}}\)), we must thus obtain estimates of \(\delta F_k\), \(\Delta \delta F_k\), \(\delta \lambda_{\text{ls}}\), and \(\Delta \delta \lambda_{\text{ls}}\). These estimates are primarily obtained through numerical methods, but it is of interest to derive formulae for a few of them in terms of the system parameters \(A\) and \(B\). Linearizing around the shifted steady state \(\delta \mathbf{y}\), we find that the linearization matrix \(A\) is shifted to \(A+\delta A\), with \(\delta A\) given by
\begin{equation}
	\label{eq:first order dA}
	\delta A_{ij} = 2 \sum_k {B_{ijk} \delta y_k}
\end{equation}
and
\begin{equation}
	\label{eq:second order dA}
	\Delta \delta A_{ij}
	= 2 \sum_k {B_{ijk} \Delta \delta y_k}
\end{equation}
Equation \eqref{eq:first order dA} also provides the rationale for the consistency constraints on the entries of array \(B\) mentioned in section \ref{sec:terminology} above, since for any arc \((j \rightarrow i) \not \in E_G\), \(\delta A_{ij} = 0\) for all perturbations \(\delta \mathbf{y}\) requires \(B_{ijk} = B_{ikj} = 0\) for all \(k\).

The shift in the system steady state to \(\delta \mathbf{y}\) under the perturbation \(\delta \mathbf{c}\) then induces a shift in the characteristic polynomial given by
\begin{equation}
	\label{eq:first order dF}
	\delta F_k = \sum_{i,j} {\frac{\partial F_k}{\partial A_{ij}} \delta A_{ij}}
\end{equation}
while under the combination of perturbations \(\delta \mathbf{y}\) and \(\Delta \mathbf{y}\),
\begin{equation}
	\label{eq:second order dF}
	\Delta \delta F_k
	= \sum_{i,j} {\left[ \begin{array}{lr}
		\frac{\partial F_k}{\partial A_{ij}} \Delta \delta A_{ij} \\
		+ \sum\limits_{(q,r) \neq (i,j)} {
			\frac{\partial^2 F_k}{\partial A_{ij} \partial A_{qr}} \Delta A_{ij} \delta A_{qr}
		}
	\end{array} \right]}
\end{equation}
where the range of the second summation in equation \eqref{eq:second order dF} excludes the pair \((i,j)\) because \(F_k\), as defined by equation \eqref{eq:feedback notation} above, is a polynomial in the entries of the matrix \(A\) in which all terms are of order 0 or 1 in any given matrix entry \(A_{ij}\) (that is, \(F_k\) is a ``multi-affine'' function of the entries of \(A\)).

Finally, we are interested also in the shift \(\delta \lambda\) in the eigenvalues \(\lambda\) under the perturbation \(\delta \mathbf{c}\). Noting that the eigenvalues \(\lambda\) must be roots of the characteristic polynomial,
\begin{equation}
	\label{eq:lambda shift equation}
	\sum_k {(F_{n-k} + \delta F_{n-k}) (\lambda + \delta \lambda)^k} = 0
\end{equation}
may be used to derive
\begin{equation}
	\label{eq:first order dlambda}
	\delta \lambda
	= \frac{-\sum\limits_k {\delta F_{n-k} \lambda^k}}{\sum\limits_k {k F_{n-k} \lambda^{k-1}}}
\end{equation}
A similar equation could be derived from equation \eqref{eq:lambda shift equation} for the mixed term \(\Delta \delta \lambda\), but in the interest of brevity, we will not consider it explicitly. Instead, we will note only that the functional dependence of the eigenvalues \(\lambda\) on the characteristic polynomial implies that the variation \(\Delta \delta \lambda\) must be a function of the variations \(\delta F_k\), \(\Delta F_k\), and \(\Delta \delta F_k\) (as well as the unperturbed eigenvalue \(\lambda\)).

\section{Sparse Matrices}
\label{sec:sparse matrices}
Consider a distribution of systems for which the structure \(G\) is a random digraph in which each arc is included with independent probability \(p_{\text{arc}}\), and for which all included arcs \((j \rightarrow i) \in E_G\) have linearization weight \(A_{ij}\) drawn from a given probability density function \(p_{\text{wt}}(x)\) independently of the weights of all other entries (arcs). Then with probability \((1-p_{\text{arc}}^k)\)  the weight \(w_K\) of any particular \(k\)-permutation term (henceforward \(k\)-term) \(K\) in equation \eqref{eq:feedback notation} is \(0\), while with probability \(p_{\text{arc}}^k\), it is distribution according to the density function
\begin{align}
	\label{eq:present k-term weight distribution}
	& p_{k\text{-wt}} \left( w_K \mid K \subset E_G \right) \\
	& = \int {p_\text{wt}\left( \left[ \prod_{i=1}^{k-1} {\frac{1}{x_i}}\right]  w_K \right)
			\left[ \prod_{i=1}^{k-1} {\frac{p_{\text{wt}}(x_i)}{|x_i|}} \right]  d^{k-1} x} \nonumber
\end{align}
since all \(k\) arcs making up \(K\) are independent and identically distributed. We further take \(p_{\text{wt}}\) symmetric about \(0\) (so that the mean of each matrix entry is \(0\)). Thus, we consider the mean-square expectation value for \(w_K\) for any \(k\)-term \(K\):
\begin{align}
	\label{eq:k-term magnitude}
	& \left\langle w_K^2 \right\rangle
	= p_{\text{arc}}^k \left\langle w_{(+,k)}^2 \right\rangle \\
	& = p_{\text{arc}}^k \int {w_K^2 \, p_{k\text{-wt}} \left( w_K \mid K \subset E_G \right) \, d \, w_K} \nonumber
\end{align}
where \(\left\langle w_{(+,k)}^2 \right\rangle\) is defined with respect to the distribution assuming all arcs in the \(k\)-term considered are present in the matrix \(A\). Note particularly that \(\left\langle w_{(+,k)}^2 \right\rangle\) does not depend on \(p_\text{arc}\). Finally, we also note that any two \(k\)-terms \(K_1\) and \(K_2\) are distinct if and only if each of \(K_1\) and \(K_2\) contains at least one arc not contained in the other: but in this case they must have zero covariance \(\left\langle\left\langle w_{K_1} w_{K_2} \right\rangle \right\rangle = 0\), since the (zero-mean) weight of each arc is independent of that of all others. Noting also that \(\langle w_K \rangle=0\) for all \(K \in \Theta_K\) --- and hence \(\langle F_k \rangle=0\) --- the mean-square value of the \(k\)\textsuperscript{th} coefficient of the characteristic polynomial is
\begin{align}
	\label{eq:char poly magnitude}
	\left\langle F_k^2 \right\rangle
	&= \sum_{K \in \Theta_k} {\left\langle w_K^2 \right\rangle}
	= \sum_{\Theta_k} {p_{\text{arc}}^k \left\langle w_{(+,k)}^2 \right\rangle } \nonumber \\
	&= \binom{n}{k} k! \, p_{\text{arc}}^k \left\langle w_{(+,k)}^2 \right\rangle
\end{align}

Now consider adding an additional \(\delta A_{qr}\) to the matrix entry \(A_{qr}\). This will result in a change \(\delta F_k\) proportional to the sum of the weights of all \(k\)-terms \(K\) containing the arc \((r \rightarrow q)\). As there are \(N_{k;qr} = \binom{n-2+\delta_{rq}}{k-2+\delta_{rq}} (k-1)!\) possible \(k\)-terms passing through the arc \((r \rightarrow q)\), each having probability \(p_{\text{arc}}^{k-1}\) of being present (given the presence of the arc \((r \rightarrow q)\)), the mean-square expectation value of \(\delta F_k\) is given by
\begin{equation}
	\label{eq:delta char poly magnitude}
	\left\langle \left[ \delta F_k \right]_{(r \rightarrow q)}^2 \right\rangle
	= N_{k;qr} p_{\text{arc}}^{k-1} \left\langle w_{(+,k-1)}^2 \right\rangle \left[ \delta A_{qr} \right]^2
\end{equation}
(Note the covariance of \(\delta F_k\) with \(F_k\) vanishes owing to the symmetry of \(p_{\text{wt}}\).) Now consider adding additional weight \(\delta A_{qr}\) and \(\Delta A_{uv}\) to the two arcs \((r \rightarrow q)\) and \((v \rightarrow u)\). Similar considerations then lead to
\begin{align}
	\label{eq:double delta char poly magnitude}
	& \left\langle \left[ \Delta \delta F_k \right]_{(v \rightarrow u), (r \rightarrow q)}^2 \right\rangle \\
	&\; \; = N_{k;qruv} p_{\text{arc}}^{k-2} \left\langle w_{(+,k-2)}^2 \right\rangle \left[ \Delta  A_{uv} \, \delta A_{qr} \right]^2 \nonumber
\end{align}
where the final term corresponds to the \(N_{k;qruv} = (1-\delta_{rv}) (1-\delta_{qu}) \binom{n-4+\delta_{qr}+\delta_{uv}+\delta_{qv}+\delta_{ru}}{k-4+\delta_{qr}+\delta_{uv}+\delta_{qv}+\delta_{ru}} (k-2)! \) possible \(k\)-terms passing through both \((r \rightarrow q)\) and \((v \rightarrow u)\).

From equations \eqref{eq:k-term magnitude}-\eqref{eq:double delta char poly magnitude}, we can thus deduce that
\begin{equation}
	\label{eq:char poly arc deriv scaling}
	\frac{\partial \sqrt{\left\langle F_k^2 \right\rangle}}{\partial A_{qr}}
	\propto p_{\text{arc}}^{\frac{1}{2} (k-1)}
	\text{, while }
	\frac{\partial^2 \sqrt{\left\langle F_k^2 \right\rangle}}{\partial A_{qr} \partial A_{uv}}
	\propto p_{\text{arc}}^{\frac{1}{2} (k-2)}
\end{equation}
from which we can conclude that for denser matrices (higher values of \(p_{\text{arc}}\)), the ratio of second- to first-derivatives of the root-mean-square expectation value of the coefficients of the characteristic polynomial \(F_k\) with respect to arc weights is generally lower than for sparser matrices.

Noting that under the variation \(A \mapsto A + \delta A + \Delta A\),
\begin{align}
	\label{eq:delta char poly deriv ratio}
	& \frac{
		\left\langle \left\langle \left[ \Delta \delta F_k \right]^2 \right\rangle \right\rangle_{\delta}
	}{
		\left\langle \left\langle \left[ \delta F_k \right]^2 \right\rangle \right\rangle_{\delta}
	} = \\
	& \frac{\sum\limits_{i,j,q,r,s,u,v,w} {\left[ \begin{array}{c}
			\frac{\partial^2 F_k}{\partial A_{ij} \partial A_{qr}}
			\frac{\partial^2 F_k}{\partial A_{su} \partial A_{vw}} \, * \\
			\Delta A_{ij} \Delta A_{su}
			\left\langle \left\langle \delta A_{qr} \, \delta A_{vw} \right\rangle \right\rangle_{\delta}
		\end{array} \right]}
	}{
		\sum\limits_{i,j,q,r} {
			\frac{\partial F_k}{\partial A_{ij}} \frac{\partial F_k}{\partial A_{qr}}
			\left\langle \left\langle \delta A_{ij} \delta A_{qr} \right\rangle \right\rangle
		}
	} \nonumber
\end{align}
we see that decreasing the relative sizes of the second derivatives of \(F_k\) compared to the first derivatives might be expected to result in decreased values of the negative term in equation \eqref{eq:uvar correlation} (with \(U=F_k\)) for the correlation \(\text{Corr}_{\delta} \left(\delta F_k + \Delta \delta F_k , \delta F_k \right)\).

At this point it should be noted that the conditions under which equations \eqref{eq:present k-term weight distribution}-\eqref{eq:delta char poly deriv ratio} were derived are quite restrictive and unrealistic if taken to describe the linearization matrices \(A\) associated with systems \(\mathbf{f}\) modeling biological systems. The simple random digraphs taken as the network structures \(G\) (especially the treatment of diagonal one-loop arcs \((i \rightarrow i)\) symmetrically with all other arcs \((r \rightarrow q)\)) and the assumption of independence of arc weights both neglect important features of real dynamic systems. In particular, under these assumptions it is almost certain that the matrix \(A\) will have some eigenvalues with positive real part, and hence will not describe the linearization of a system \(\mathbf{f}\) about a stable steady state.

We do not attempt to address these points analytically, but instead shift our attention to numerical investigation of specific systems with more complex (and realistic) features. This also allows us to consider another dynamic system property --- the distance \(\text{Re}(\lambda_\text{ls})\) from the set of eigenvalues of \(A\) to the imaginary axis in the complex plane (here designated the eigenvalue stability) --- which is of more immediate use in determining system stability.

\section{Arc-Derivative Ratios for Stabilized Matrices}
\label{sec:numerical arc analysis}
We first consider the impact of requiring system stability --- i.e., all eigenvalues of the linearization matrix \(A\) must be in the left half of the complex plane --- on the ratio of second derivatives of the determinant or the eigenvalue stability to the first derivatives with respect to the arc weights \(A_{ij}\). Here we enforce this requirement on each matrix \(A\) by subtracting \((\text{Re}(\lambda_{\text{ls}})+1)I\) from \(A\), where \(\lambda_{\text{ls}}\) is (one of) the largest real-part eigenvalue(s) of \(A\), so that the new version of \(A\) will always have eigenvalue stability -1. (From the point of view of the network \(G\) corresponding to the matrix \(A\), this operation consists of adding all one-loop arcs to \(G\) with weight \(-(\text{Re}(\lambda_{\text{ls}})+1)\).)

\begin{figure*}[!t]
\centering
\fbox{\begin{minipage}{\textwidth}
	\includegraphics[width=\textwidth]{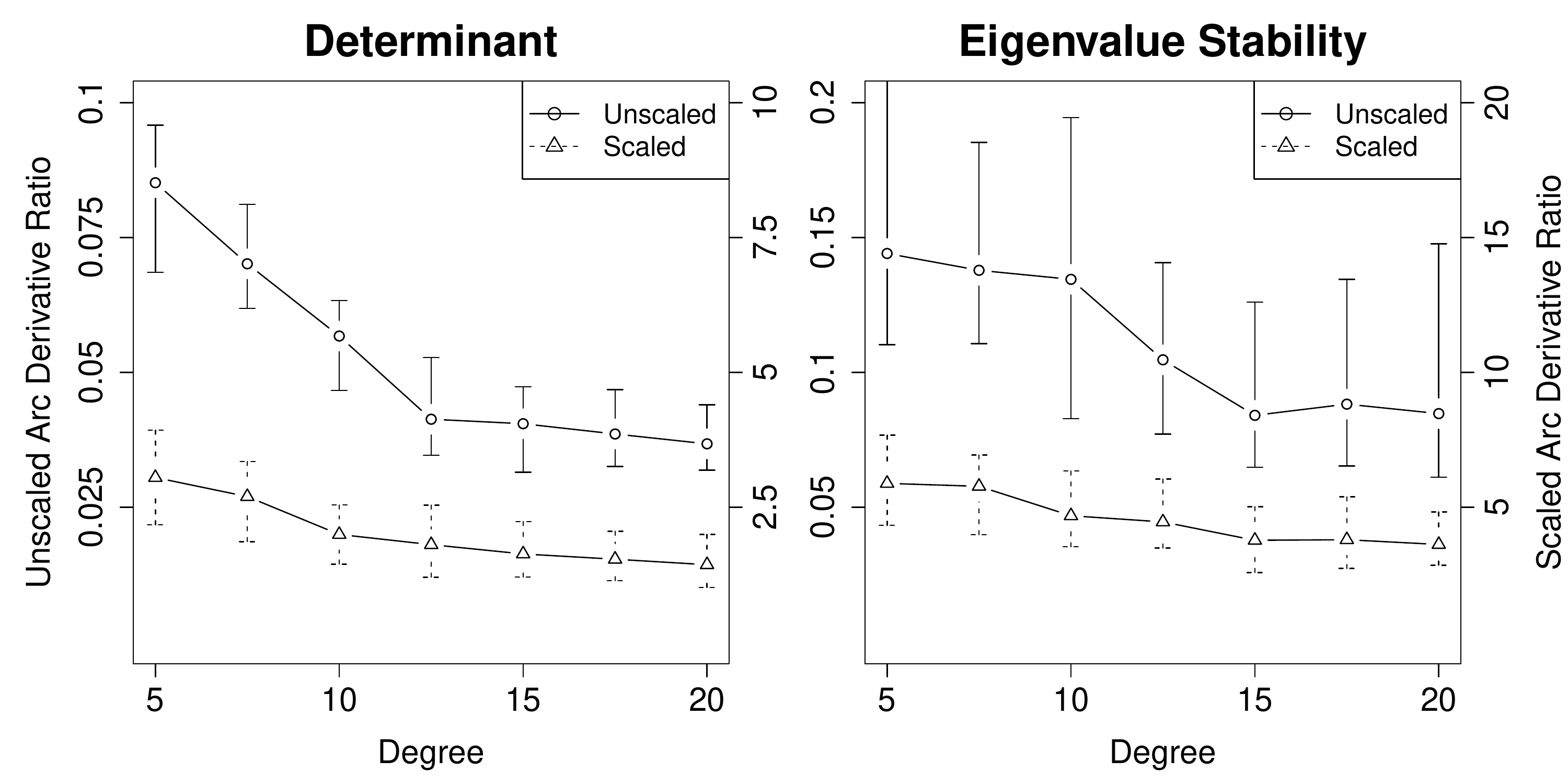}
	\caption{Median values of the ratio described by equation \eqref{eq:basic deriv ratio} from randomly generated matrices (100 at each point) with random digraph structure of varying average node degree. Ratios are presented unscaled or scaled according to equations \eqref{eq:first order uvar scaling}-\eqref{eq:second order uvar scaling}, as indicated. Y-axis labels for unscaled ratios on left, scaled to right. Error bars represent first and third quartiles.}
	\label{fig:arc deriv varying size}
\end{minipage}}
\end{figure*}

In order to characterize the magnitudes of second- relative to first-derivatives of both the determinants \(\text{det}(A) = (-1)^{n-1} F_n\) and the eigenvalue stabilities \(\text{Re}(\lambda_{\text{ls}})\) with respect to arc weights \(A_{ij}\) for a given matrix \(A\) with randomly constructed network structure \(G\), we numerically evaluated the 25 first derivatives and 625 second derivatives of both metrics with respect to a set \(E_{\text{ptb}}\) containing 25 arcs randomly chosen from the arc set \(E_G\). The root-mean-square (rms) average of the 625 second derivative terms was then divided by the rms average of the 25 first derivative terms to yield a single value
\begin{equation}
	\label{eq:basic deriv ratio}
	\frac{\sqrt{
		\frac{1}{\left| E_{\text{ptb}} \right|^2}
		\sum\limits_{(j \rightarrow i), (r \rightarrow q) \in E_{\text{ptb}}} {
			\left[ \frac{\partial^2 U}{\partial A_{ij} \partial A_{qr}} \right]^2
		}
	}}{\sqrt{
		\frac{1}{\left| E_{\text{ptb}} \right|}
		\sum\limits_{(j \rightarrow i) \in E_{\text{ptb}}} {
			\left[ \frac{\partial U}{\partial A_{ij}} \right]^2
		}
	}}
\end{equation}
of this ratio for each matrix \(A\) for each of the two metrics (determinant and eigenvalue stability).

As mentioned in section \ref{sec:uvar} above, when comparing the response of systems with widely varying structures to parametric perturbation, a measure of the perturbation size is required. For our purposes, it is natural to use the induced change to the eigenvalue stability as an indicator of the relative magnitude of a given perturbation, especially given that we have already normalized all of our stabilized systems to have the same base eigenvalue stability value of -1. In the discussion to follow, we thus also present results for (rms averages of)  derivatives with respect to the scaled arc weights
\begin{equation}
	\tilde{A}_{ij}
	= \frac{1}{\epsilon} \sqrt{\left\langle \left[
		\Delta \text{Re}(\lambda_{\text{ls}})
	\right]^2 \right\rangle_{E_{\text{ptb}}}} \, A_{ij}
\end{equation}
where
\begin{equation}
	\label{eq:delta eig stab estimate}
	\left\langle \left[
		\Delta \text{Re}(\lambda_{\text{ls}}) 
	\right]^2 \right\rangle_{E_{\text{ptb}}} \\
	= \frac{\epsilon^2}{\left| E_{\text{ptb}} \right|} \sum_{(j \rightarrow i) \in E_{\text{ptb}}} {\left[
		\frac{\partial \text{Re}(\lambda_{\text{ls}})}{\partial A_{ij}}
	\right]^2}
\end{equation}
is approximately equal to the mean-square variation of \(\text{Re}(\lambda_\text{ls})\) if the weights of the arcs in \(E_\text{ptb}\) are subjected to independent Gaussian perturbations of equal (small) variance \(\frac{\epsilon^2}{|E_\text{ptb}|}\). The resulting scaled derivatives are then
\begin{equation}
	\label{eq:first order uvar scaling}
	\frac{\partial U}{\partial \tilde{A}_{ij}}
	= \frac{1}{\epsilon} \left\langle \left[
		\Delta \text{Re}(\lambda_{\text{ls}}) 
	\right]^2 \right\rangle_{E_{\text{ptb}}}^{-\frac{1}{2}} \,
	\frac{\partial U}{\partial A_{ij}}
\end{equation}
and
\begin{equation}
	\label{eq:second order uvar scaling}
	\frac{\partial^2 U}{\partial \tilde{A}_{ij} \partial \tilde{A}_{qr}}
	= \frac{1}{\epsilon^2} \left\langle \left[
		\Delta \text{Re}(\lambda_{\text{ls}}) 
	\right]^2 \right\rangle_{E_{\text{ptb}}}^{-1} \,
	\frac{\partial^2 U}{\partial A_{ij} \partial A_{qr}}
\end{equation}
where \(U\) is either the determinant \(\pm F_n\) or the eigenvalue stability Re\((\lambda_{\text{ls}})\). This scaling is specifically indicated where it has been applied. Clearly, for \(U=\text{Re}(\lambda_{\text{ls}})\), one result of such scaling will be to bring the rms average of the first derivatives to unity.

Figure \ref{fig:arc deriv varying size} then shows the first, second (median), and third quartiles of these ratios calculated from populations of stabilized random digraphic-structured matrices with varying arc density (100 matrices, each with a different structure \(G\) and \(n=100\), were generated for each arc density grouping). Both the unscaled and scaled data exhibit the expected trend of decreasing rms magnitudes of second derivatives compared to first derivatives for metric \(U\) chosen to be either determinant or eigenvalue stability.

While these results suggest that that the qualitative effects of network density suggested by equations \eqref{eq:char poly arc deriv scaling} still hold for stabilized random digraphical systems with varying arc densities (and perturbation sensitivities), we have not yet considered more complex structural variation. We also have not yet considered correlation under variation  of the offset parameters originally introduced in section \ref{sec:uvar}, as opposed to the derivatives with respect to individual arc weights considered in this section. The behavior of these correlations (for random digraphical systems of varying arc density, for scale-free systems characterized by large node degree heterogeneity, and for highly clustered small-world systems) is investigated in the next section.

\section{Correlation Susceptibility in Perturbed Dynamic Systems}
\label{sec:numerical analysis pert}

In order to study the effects of network structure on signal integration in non-linear dynamic systems described by equation \eqref{eq:simplified general system}, it is necessary to consider not only the network structure \(G\) and the linearization matrix \(A\), but also the quadratic terms \(B\). For the purposes of this paper, random systems were constructed by first generating a random network \(G\) (according to one of the random digraph generation algorithms described in appendix \ref{sec:network generation}), followed by progressively building up \(A\) and \(B\) as follows.

For each arc \((j \rightarrow i) \in E_G\), we generate a random ``rate constant'' \(r_{ij}\) from a Gaussian distribution of vanishing mean and unit variance independent of all other system parameters and reset
\begin{align}
	\label{eq:system generation}
	A_{ij} &\mapsto A_{ij} + r_{ij} \\
	B_{iji} &\mapsto B_{iji} + \frac{1}{2} r_{ij} \nonumber \\
	B_{iij} &\mapsto B_{iij} + \frac{1}{2} r_{ij} \nonumber
\end{align}
The contribution of arc \((j \rightarrow i)\) to the system described by equation \eqref{eq:system generation} can be taken to represent a reaction
\begin{equation}
	\label{eq:rxn notation}
	\text{sp}_i + \text{sp}_j \rightarrow \left[ 1 + \text{sign}(r_{ij}) \right] \text{sp}_i + \text{sp}_j
\end{equation}
with mass-action kinetics (offset by a reaction \(\text{sp}_i \rightarrow \left[ 1 - \text{sign}(r_{ij}) \right] \text{sp}_i \) with equal magnitude rate constant), where \(x_i\) measures the standardized deviation of the population of species \(i\) from it steady state value.

After all arcs in \(E_G\) have been accounted for, the system \(S\) is stabilized by adding a multiple of the identity matrix to \(A\) so as to bring \(\text{Re}(\lambda_{\text{ls}})\) to -1, as described in section \ref{sec:numerical arc analysis} above (thereby including all one-loops \((i \rightarrow i)\) in the final structure \(G\)).

We then investigate the effects of perturbations \(\delta \mathbf{c}\) and \(\Delta \mathbf{c}\), as defined by equation \eqref{eq:pert signal}, on the determinant \(\pm F_n\) and eigenvalue stability \(\text{Re}(\lambda_\text{ls})\). Taking \(\Delta \mathbf{c}\) to be fixed and \(\delta \mathbf{c}\) to be a random vector with 
covariance matrix \(\sigma^2 I\) (and vanishing mean), and defining for any function \(U\) of the parameters \((A,B,\mathbf{c})\) of the system about its steady state,
\begin{equation}
	\label{u vector definition}
	u_i = \frac{\partial U}{\partial c_i}
\end{equation}
and
\begin{equation}
	\label{W matrix definition}
	W_{ij} = \frac{\partial^2 U}{\partial c_i \partial c_j} = \frac{\partial^2 U}{\partial c_j \partial c_i}
\end{equation}
the variances and covariances of the changes \(\delta U\) and \(\Delta \delta U\) to the arbitrary function \(U\) resulting from the random perturbation \(\delta \mathbf{c}\) in the presence of the fixed perturbation \(\Delta \mathbf{c}\) may be derived from
\begin{align}
	\label{eq:dU variance}
	\left\langle \left\langle
		\delta U \, \delta U
	\right\rangle \right\rangle_{\delta}
	&= \sum_{i,j} {
		\frac{\partial U}{\partial c_i}
		\frac{\partial U}{\partial c_j}
		\left\langle \left\langle
			\delta c_i \, \delta c_j
		\right\rangle \right\rangle_{\delta}
	} \\
	&= \sigma^2 \mathbf{u} \cdot \mathbf{u} \nonumber \\
	\left\langle \left\langle
		\delta U \, \Delta \delta U
	\right\rangle \right\rangle_{\delta}
	&= \sigma^2 \sum_{i,j} {
		\frac{\partial U}{\partial c_i}
		\frac{\partial^2 U}{\partial c_j \partial c_i}
		\Delta c_j
	} \nonumber \\
	&= \sigma^2 \mathbf{u}^{\text{T}} W \Delta \mathbf{c} \nonumber \\
	\left\langle \left\langle
		\Delta \delta U \, \Delta \delta U
	\right\rangle \right\rangle_{\delta}
	&= \sigma^2 \sum_{i,j,k} {
		\frac{\partial^2 U}{\partial c_j \partial c_i}
		\frac{\partial^2 U}{\partial c_k \partial c_i}
		\Delta c_j \Delta c_k
	} \nonumber \\
	&= \sigma^2 \left( W \Delta \mathbf{c} \right) \cdot \left( W \Delta \mathbf{c} \right) \nonumber
\end{align}

As discussed in section \ref{sec:uvar} above, we use the correlation equation \eqref{eq:uvar correlation} as a proxy metric to investigate the propensity of a system towards signal integration. For the reasons discussed in section \ref{sec:numerical arc analysis}, we consider eigenvalue stability-normalized perturbations satisfying
\begin{equation}
	\label{eq:pert normalization constraint}
	\left\langle \left[ \delta \text{Re} \left( \lambda_\text{ls} \right) \right]^2 \right\rangle_\delta = \phi^2
\end{equation}
where \(\phi\) is a constant across all systems considered. Taking \(U = \text{Re}( \lambda_\text{ls} ) \) and employing equation \eqref{eq:dU variance}, obtain
\begin{equation}
	\label{eq:pert normalization constraint 2}
	\left\langle \left[ \delta \text{Re} \left( \lambda_\text{ls} \right) \right]^2 \right\rangle_\delta
	= \sigma^2 \sum_i {\left[ \frac{\partial \text{Re} \left( \lambda_\text{ls} \right)}{\partial c_i} \right]^2}
\end{equation}
thus implying
\begin{equation}
	\label{eq:pert normalization constraint 3}
	\sigma^2 = \left(
		\sum_i {\left[ \frac{\partial \text{Re} \left( \lambda_\text{ls} \right)}{\partial c_i} \right]^2}
	\right)^{-1} \phi^2
	= \frac{\phi^2}{\mathbf{u}_\text{es} \cdot \mathbf{u}_\text{es}}
\end{equation}
where \(\mathbf{u}_\text{es}\) is the gradient vector of the eigenvalue stability \(U_\text{es} = \text{Re}(\lambda_\text{ls})\) with respect to the dynamic offset vector \(\mathbf{c}\). Similar normalization of the interacting perturbation \(\Delta \mathbf{c}\) then requires:
\begin{equation}
	\label{eq:pert normalization constraint 4}
	\left\langle \left\langle \Delta c_i \, \Delta c_j \right\rangle \right\rangle_\Delta
	= \frac{\Phi^2}{\mathbf{u}_\text{es} \cdot \mathbf{u}_\text{es}} \delta_{ij}
\end{equation}
where \(\Phi\) is, again, a system-independent constant, and we now consider \(\Delta \mathbf{c}\) to be drawn from a probability distribution with vanishing mean \(\langle \Delta \mathbf{c} \rangle = \mathbf{0}\) and covariance matrix proportional to the identity (similar to \(\delta \mathbf{c}\) above). Equations \eqref{eq:dU variance} - \eqref{eq:pert normalization constraint 4} together with equation \eqref{eq:uvar correlation} imply
\begin{equation}
	\label{eq:uvar correlation detailed}
	\text{Corr}_\delta \left(
		\delta U , \delta U + \Delta \delta U
	\right)
	= 1 - \Phi^2 \Upsilon_{U;\Delta} + \text{H.O.T.}
\end{equation}
 where the \(\Delta\)-correlation susceptibility \(\Upsilon_{U;\Delta}\) is given by
\begin{equation}
	\label{eq:uvar correlation susceptibility}
	\Upsilon_{U;\Delta}
	= \frac{1}{2 \Phi^2} \left[
		\frac{\left( W \Delta \mathbf{c} \right) \cdot \left( W \Delta \mathbf{c} \right)}{\mathbf{u} \cdot \mathbf{u}}
		- \left( \frac{\mathbf{u} \cdot W \Delta \mathbf{c}}{\mathbf{u} \cdot \mathbf{u}} \right)^2
	\right]
\end{equation}
The quantity \(\Phi^2 \Upsilon_{U;\Delta}\) represents the degree to which \(\text{Corr}_\delta (\delta U, \delta U + \Delta \delta U)\) is reduced in the context of the interacting perturbation \(\Delta \mathbf{c}\) normalized to produce a first-order change \(\Delta \text{Re}(\lambda_\text{ls})\) of mean-square magnitude
\begin{equation}
	\label{eq:delta es magnitude}
	\left\langle \left[ \Delta \text{Re} \left( \lambda_\text{ls} \right) \right]^2 \right\rangle_\Delta = \Phi^2
\end{equation}

\begin{figure*}[t]
\centering
\fbox{\begin{minipage}{\textwidth}
	\includegraphics[width=\textwidth]{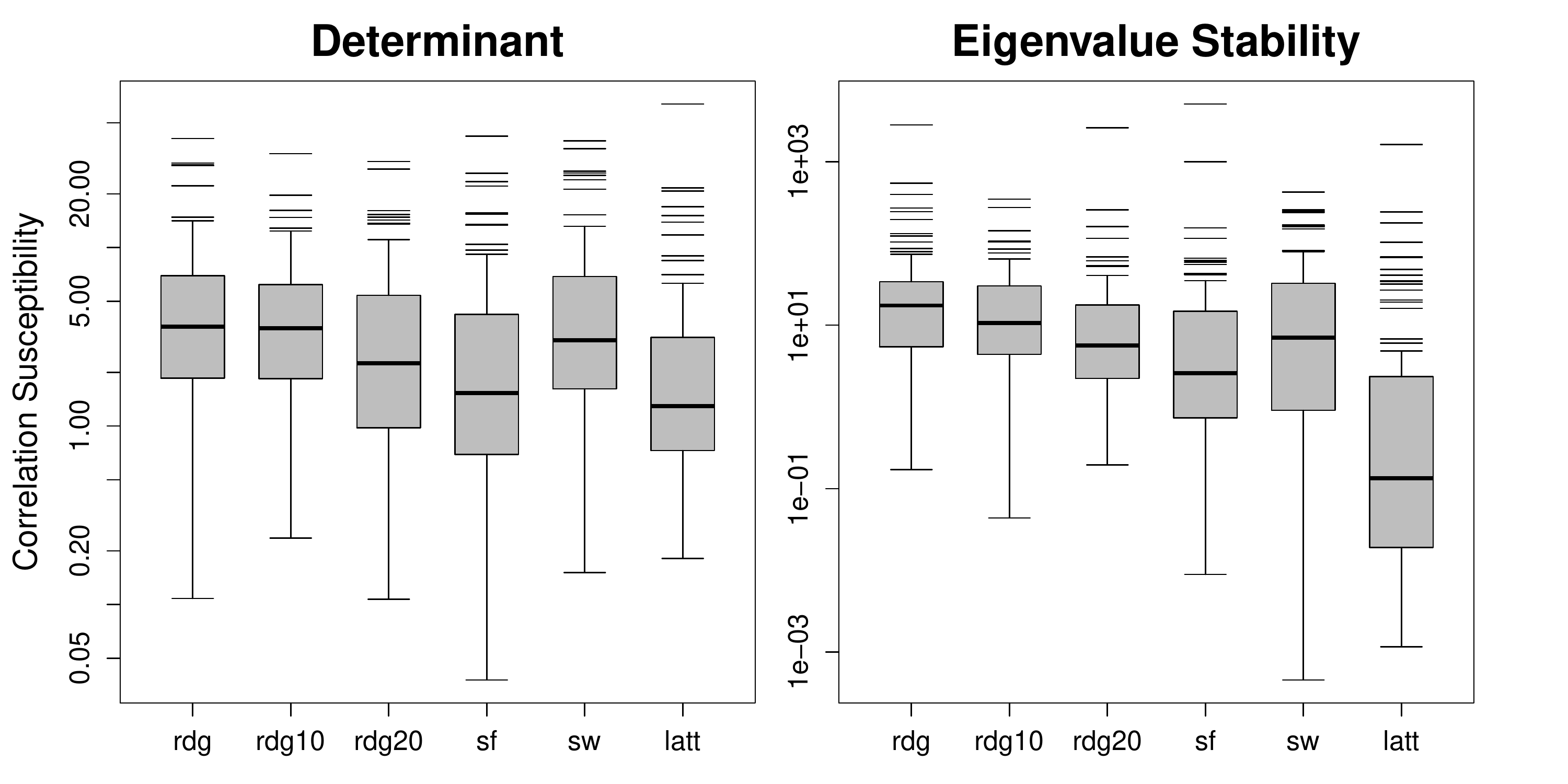}
	\caption{Boxplots (1.5 IQR whiskers) of determinant and eigenvalue stability correlation susceptibilities described by equation \eqref{eq:average uvar correlation susceptibility} for varying types of network structure. Random digraphical systems of average node degree 6, 10, and 20 indicated by \textbf{rdg}, \textbf{rdg10}, and \textbf{rdg20}, respectively, while scale-free, small-world, and ring-lattice digraphs (all of average node degree 6) indicated by \textbf{sf}, \textbf{sw}, and \textbf{latt}. \(n=100\) nodes for all systems.}
	\label{fig:correlation susceptibility bxp}
\end{minipage}}
\end{figure*}

The \(\Delta\)-correlation susceptibility \(\Upsilon_{U;\Delta}\) depends not only on the particular system under consideration but also on the choice of perturbation \(\Delta \mathbf{c}\). As we are interested ultimately in characterizing the propensity of systems themselves toward signal integration or independence, we now further average over the interacting signal \(\Delta \mathbf{c}\) to define the (average) correlation susceptibility,
\begin{equation}
	\label{eq:average uvar correlation susceptibility}
	\Upsilon_U = \left\langle \Upsilon_{U;\Delta} \right\rangle_\Delta
\end{equation}
\(\Upsilon_U\) (equation \eqref{eq:average uvar correlation susceptibility}) is henceforward referred to as simply the correlation susceptibility.

Figure \ref{fig:correlation susceptibility bxp} provides boxplots of the correlation susceptibilities \(\Upsilon_U\) for random digraph-, scale-free-, small-world-, and lattice-structured systems (see appendix \ref{sec:network generation} for details regarding directed network structures) generated as described above. ANOVA models of the logged correlation susceptibilities against the six network groupings shown in figure \ref{fig:correlation susceptibility bxp} indicate significant effects (determinant \(F_{5, 594}=9.68\) with \(p<10^{-8}\) and \(R^2=0.075\), eigenvalue stability \(F_{5, 594}=40.41\) with \(p<10^{-16}\) and \(R^2=0.25\)).

As expected, figure \ref{fig:correlation susceptibility bxp} shows that the random digraph systems (\textbf{rdg}, \textbf{rdg10}, and \textbf{rdg20}) have progressively lower determinant- and eigenvalue stability-correlation susceptibilities as the arc density increases. We can now also compare the more complex scale-free (\textbf{sf}) and small-world (\textbf{sw} and \textbf{latt}) structured systems, however: both of these forms of structure result in lowered correlation susceptibility (determinant and eigenvalue stability) when compared with random digraphical systems of the same arc density. In section \ref{sec:structural metrics}, we pursue structural metrics on the basis of the arc density arguments made in section \ref{sec:sparse matrices} designed to predict correlation susceptibility.

\section{Relevant Structural Metrics}
\label{sec:structural metrics}

In sections \ref{sec:sparse matrices} - \ref{sec:numerical analysis pert}, it has been argued that increasing network arc density leads to decreased values of our proxy for signal integration, correlation susceptibility (both of the determinant and of the eigenvalue stability). One might expect, however, that not all arcs have an equal degree of influence on correlation susceptibility; both the weight \(A_{ji}\) of an arc \((i \rightarrow j)\) and the location of its termini \(i\) and \(j\) in the network suggest themselves as important factors to consider. Thus we seek a scalar metric derived from the linearization matrix \(A\) in the form of a sum of normalized arc weights, with the normalization factor hopefully capturing some of the influence of arc locality.

It is not immediately obvious how to go about constructing such a normalization factor from the linearization matrix \(A\). For the sake of clarity and parsimony, we begin by postulating that, since system stability is largely a function of the least-stable eigenvalue(s) \(\lambda_\text{ls}\)  of \(A\), the eigenvector(s) associated with \(\lambda_\text{ls}\) may offer a useful first indication to the relative influence of the various system nodes in questions of stability.

\begin{figure*}[!t]
\centering
\fbox{\begin{minipage}{\textwidth}
	\includegraphics[width=\textwidth]{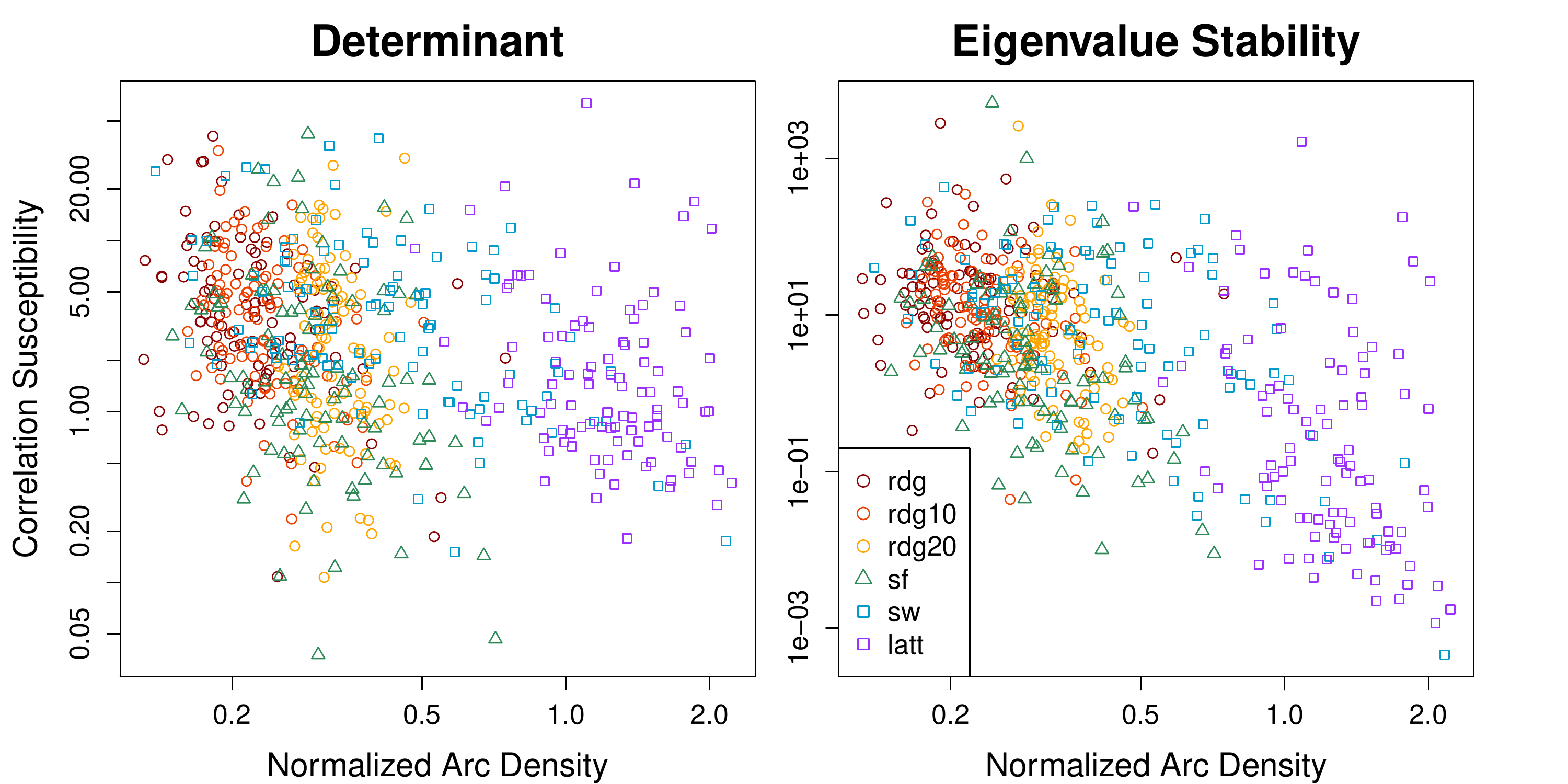}
	\caption{Determinant and eigenvalue stability correlation stabilities (equation \eqref{eq:average uvar correlation susceptibility}) versus normalized arc density (equation \eqref{eq:normalized arc density}) for varying types of network structure. Random digraphical systems of average node degree 6, 10, and 20 indicated by \textbf{rdg}, \textbf{rdg10}, and \textbf{rdg20}, respectively, while scale-free, small-world, and ring-lattice digraphs (all of average node degree 6) indicated by \textbf{sf}, \textbf{sw}, and \textbf{latt}. \(n=100\) nodes for all systems.}
	\label{fig:correlation susceptibility vs norm arc density}
\end{minipage}}
\end{figure*}

Let \(\mathbf{v}\) be (one of) the (generally complex) least-stable eigenvector(s) of the linearization matrix \(A\), normalized so that \(\mathbf{v} \cdot \mathbf{v} = 1\). We here define the \emph{normalized arc density \(\rho_A\)} by
\begin{equation}
	\label{eq:normalized arc density}
	\rho_A = \sum_{(i \rightarrow j) \in E_G} {\left| A_{ji} \right| \left|v_i \right|^2 \left| v_j \right|^2}
\end{equation}
For systems constructed as described in section \ref{sec:numerical analysis pert}, there will almost always be either exactly one (if \(\lambda_\text{ls}\) is real) or two (if \(\lambda_\text{ls}\) is complex) least stable eigenvectors. When \(\lambda_\text{ls}\) is complex, the two least stable eigenvectors will be complex conjugates of each other; this implies that the value of \(\rho_A\) as defined by equation \eqref{eq:normalized arc density} will be independent of which eigenvector is chosen.

\begin{table*}[b!]
\centering
\begin{tabular}{rcccccccc}
	\toprule
	& \(\Upsilon_\text{det}\) & \(\Upsilon_\text{es}\) & \(\rho_A\) & \(\Psi_A\) \\ 
	\midrule
	rdg & 3.6 \(\pm\) 3.1 & 17 \(\pm\) 19 & 0.22 \(\pm\) 0.05 & 0.015 \(\pm\) 0.010 \\ 
	rdg10 & 3.5 \(\pm\) 3.0 & 11 \(\pm\) 13 & 0.23 \(\pm\) 0.05 & 0.012 \(\pm\) 0.009 \\ 
	rdg20 & 2.2 \(\pm\) 2.4 & 5.7 \(\pm\) 7.6 & 0.32 \(\pm\) 0.05 & 0.012 \(\pm\) 0.008 \\ 
	sf & 1.5 \(\pm\) 1.7 & 2.6 \(\pm\) 3.6 & 0.29 \(\pm\) 0.08 & 0.018 \(\pm\) 0.016 \\ 
	sw & 3.0 \(\pm\) 3.0 & 7.0 \(\pm\) 10 & 0.40 \(\pm\) 0.21 & 0.016 \(\pm\) 0.013 \\ 
	latt & 1.3 \(\pm\) 1.1 & 0.13 \(\pm\) 0.20 & 1.2 \(\pm\) 0.39 & 0.030 \(\pm\) 0.026 \\ 
	\bottomrule \end{tabular}
	\caption{Median \(\pm\) MAD values for determinant and eigenvalue stability correlation susceptibilities (equation \eqref{eq:average uvar correlation susceptibility}), normalized arc densities (equation \eqref{eq:normalized arc density}), and eigenvector contractions (equation \eqref{eq:eigenvector contraction}) for the systems used to generate figures \ref{fig:correlation susceptibility bxp} - \ref{fig:correlation susceptibility vs norm arc density}.}
	\label{table:network metrics}
\end{table*}

Figure \ref{fig:correlation susceptibility vs norm arc density} shows the association of the correlation susceptibility and the normalized arc density (\(R^2 \left(\text{log}(\Upsilon_\text{det}),\text{log}(\rho_A) \right)=0.10\), \(R^2 \left(\text{log}(\Upsilon_\text{es}),\text{log}(\rho_A) \right)=0.35\)). Considering only \(R^2\), the normalized arc density thus appears to explain more of the influence of network structure on correlation susceptibility than the ANOVA (section \ref{sec:numerical analysis pert}) on the network groupings themselves (this is not a completely fair comparison, however, since the normalized arc density includes more detailed information arising from the arc weights \(A_{ij}\)). Additionally, the normalized arc density is capable of differentiating network structures with the same ``raw'' arc density: consulting table \ref{table:network metrics}, note that scale-free and, especially, small-world networks generally have larger normalized arc densities than random digraphs of the same average degree.

Equation \eqref{eq:normalized arc density} considers only the magnitudes of the components of the least-stable eigenvector(s) \(\mathbf{v}\). Considering equation \eqref{eq:feedback notation}, however, we see that the characteristic polynomial (and hence ultimately the eigenvalues) of the linearization matrix \(A\) depend on sums of products of the weights of arcs making up permutation \(k\)-terms. If, for example, the weights \(A_{12}\) and \(A_{21}\) change as the result of a perturbation by \(\delta A_{12}\) and \(\delta A_{21}\), the resulting change to the weight of the \(k\)-term \(K=(1 \leftrightarrow 2)\) is
\begin{equation}
	\label{eq:k-term dA phase dependence}
	\delta w_K = A_{12} \delta A_{21} + \delta A_{12} A_{21}
\end{equation}
This quantity depends on the relative phases of the arc weight perturbations, which will in turn depend on the relative phases of the perturbations to the components of the system steady state vector. Thus, we might suspect that the phases, as well as the magnitudes, of the components of the least-stable eigenvector(s), contain information relevant to the correlation susceptibility of a system.

\begin{figure*}[!t]
\centering
\fbox{\begin{minipage}{\textwidth}
	\includegraphics[width=\textwidth]{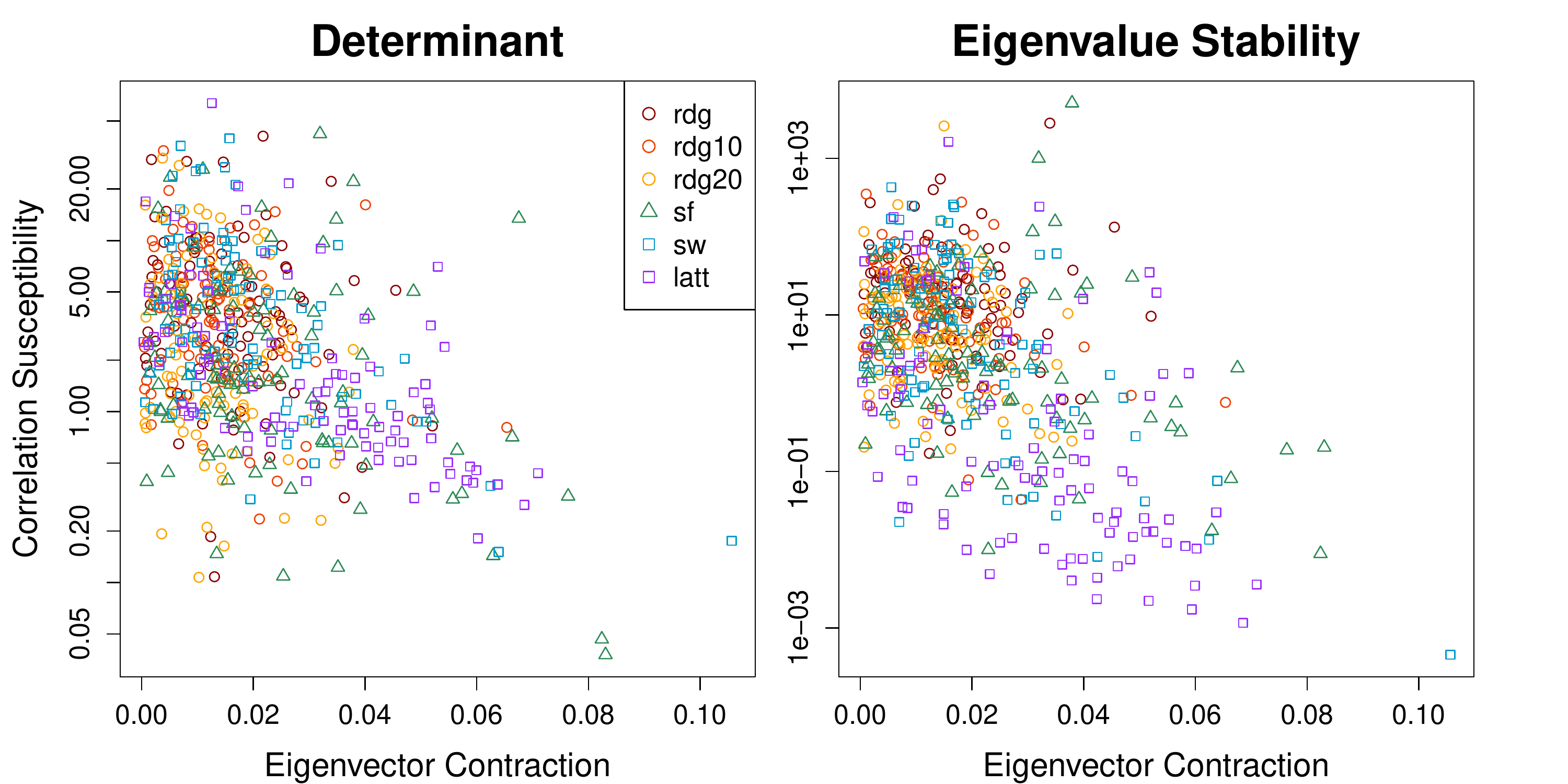}
	\caption{Determinant and eigenvalue stability correlation susceptibilities (equation \eqref{eq:average uvar correlation susceptibility}) versus eigenvector contraction (equation \eqref{eq:eigenvector contraction}) for varying types of network structure. Random digraphical systems of average node degree 6, 10, and 20 indicated by \textbf{rdg}, \textbf{rdg10}, and \textbf{rdg20}, respectively, while scale-free, small-world, and ring-lattice digraphs (all of average node degree 6) indicated by \textbf{sf}, \textbf{sw}, and \textbf{latt}. \(n=100\) nodes for all systems.}
	\label{fig:correlation susceptibility vs eigenvector contraction}
\end{minipage}}
\end{figure*}

Thus, letting \(\Xi = [A^{-1} (A^{-1})^\text{T}]\), define the \emph{eigenvector contraction} by 
\begin{equation}
	\label{eq:eigenvector contraction}
	\Psi_A = \left| \frac{\sum\limits_i {\Xi_{ii} v_i}}{\text{Tr}(\Xi)} \right|
\end{equation}
(Equation \eqref{eq:eigenvector contraction} is again generally independent of the choice of least-stable eigenvector when \(\lambda_\text{ls}\) is complex, since the absolute value is invariant with respect to complex conjugation of its argument.) Note that \(\Psi_A\) is a weighted sum of the (phased) components of the least stable eigenvector \(\mathbf{v}\), with the weightings provided by the diagonal elements of the matrix \(\Xi\). This choice of weighting was motivated by the observation that
\begin{equation}
	\label{eq:node variances}
	\left\langle \left\langle \delta y_i \, \delta y_j \right\rangle \right\rangle_\delta
	= \sigma^2 \sum_k {A_{ik}^{-1} A_{jk}^{-1}}
	= \sigma^2 \Xi_{ij}
\end{equation}
(derived from \(\text{Cov}(\delta \mathbf{c})=\sigma^2 I\) and equation \eqref{eq:first order dy}). That is, we weight more heavily those components of the least stable eigenvector corresponding to nodes whose components in the steady state vector \(\mathbf{y}\) undergo larger variance under the perturbations \(\delta \mathbf{c}\).

Figure \ref{fig:correlation susceptibility vs eigenvector contraction} plots the correlation susceptibility against the eigenvector contraction metric (\(R^2 \left(\text{log}(\Upsilon_\text{det}), \Psi_A \right)=0.19\), \(R^2 \left(\text{log}(\Upsilon_\text{es}), \Psi_A \right)=0.27\)). Once again, the \(R^2\) values of this metric are larger than those of the ANOVA from section \ref{sec:numerical analysis pert} on the network groupings --- especially, in this case, with regard to the determinant correlation susceptibility. Compared to the normalized arc density, however, the eigenvector contraction has a relatively large within-network-group MAD-to-median ratio, indicating that this factor perhaps helps to explain more of the within-group variance in correlation susceptibility.

\begin{figure*}[!t]
\centering
\fbox{\begin{minipage}{\textwidth}
	\includegraphics[width=\textwidth]{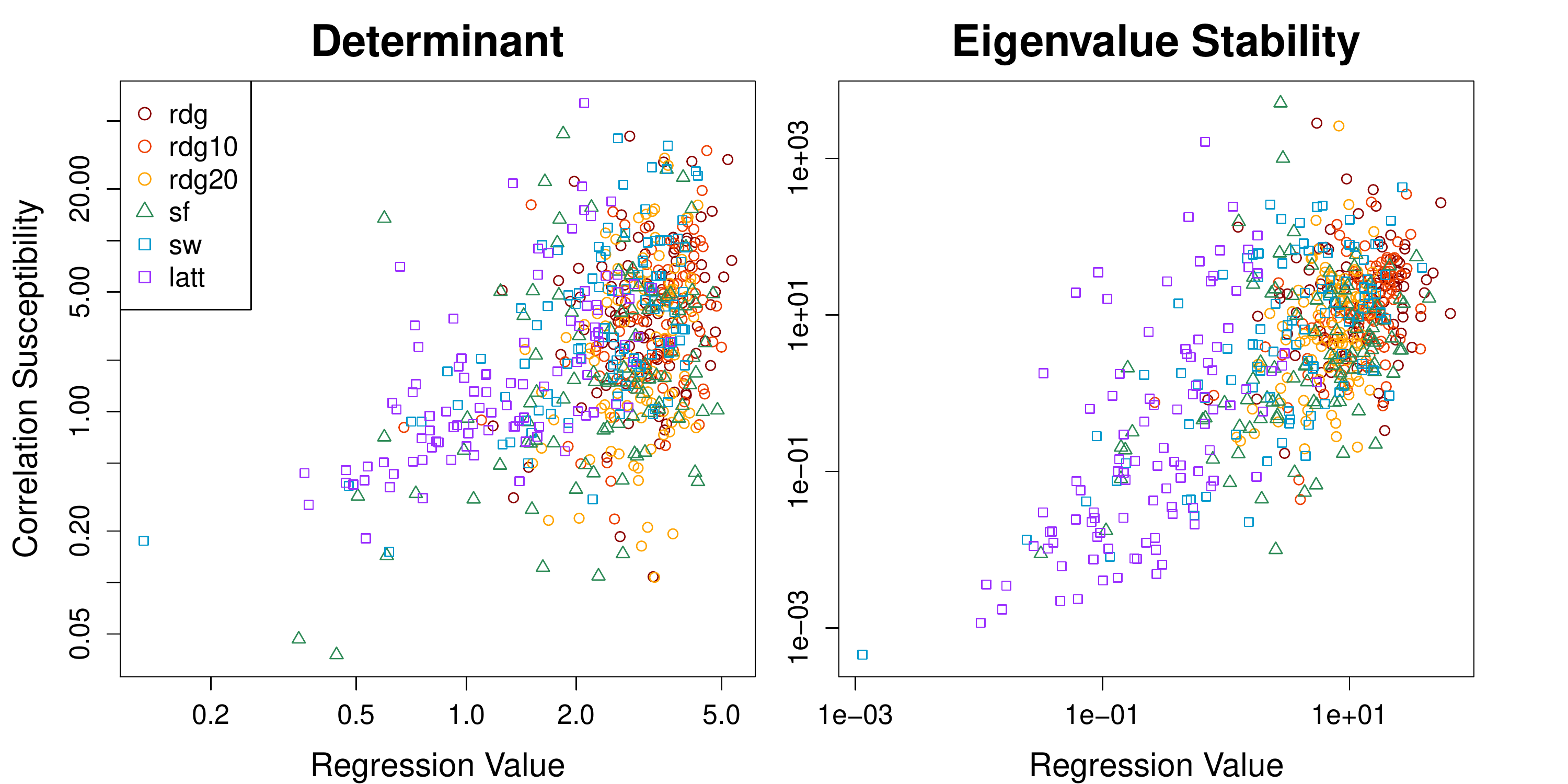}
	\caption{Determinant and eigenvalue stability correlation susceptibilities (equation \eqref{eq:uvar correlation susceptibility}) versus regression value (\textbf{equation XXX}) for varying types of network structure. Random digraphical systems of average node degree 6, 10, and 20 indicated by \textbf{rdg}, \textbf{rdg10}, and \textbf{rdg20}, respectively, while scale-free, small-world, and ring-lattice digraphs (all of average node degree 6) indicated by \textbf{sf}, \textbf{sw}, and \textbf{latt}. \(n=100\) nodes for all systems.}
	\label{fig:correlation susceptibility regression}
\end{minipage}}
\end{figure*}

The correlation of the eigenvector contraction and the two correlation susceptibilities is negative, similar to the relationship between normalized arc density and correlation susceptibilities. Following the arguments made in section \ref{sec:sparse matrices}, one might consider the relative impact of in-phase- versus out-of-phase-perturbations to the arc weights (originating from perturbations \(\delta \mathbf{y}\) to the system steady state) on the first- as compared to the second-derivatives of the \(k\)-terms composing the characteristic polynomial \(F_k\). For the sake of brevity, this paper does not continue on this path, noting only the empirical correlation between contraction and susceptibility.

\begin{table}
\centering
\begin{tabular}{cccc}
	\toprule 
	&  Estimate & Std Error & \(\text{Pr}(>|t|)\) \\ 
	\midrule
	\(\beta_{\text{log}(\rho_A)}^\text{det}\) & -0.171 & 0.040 & 1.87E-5 \\ 
	\(\beta_{\Psi_A}^\text{det}\) & -0.363 & 0.040 & 6.66E-19 \\
	\midrule
	\(\beta_{\text{log}(\rho_A)}^\text{es}\) & -0.459 & 0.033 & 3.53E-38 \\ 
	\(\beta_{\Psi_A}^\text{es}\) & -0.339 & 0.033 & 7.67E-23 \\
	\bottomrule
\end{tabular}
\caption{Standardized regression coefficients for models described by equation \eqref{eq:correlation susceptibility regression}. The resulting fits have \(R_\text{det}^2=0.21\) and \(R_\text{es}^2=0.45\).} 
\label{table:correlation susceptibility regression}
\end{table}

Table \ref{table:correlation susceptibility regression} provides the results of regression fits
\begin{equation}
	\label{eq:correlation susceptibility regression}
	\widetilde{\text{log}(\Upsilon_U)}
	= \beta_{\text{log}(\rho_A)}^U \widetilde{\text{log}(\rho_A)}
	+ \beta_{\Psi_A}^U \widetilde{\Psi}_A
\end{equation}
for \(U\) either determinant or eigenvalue stability; here we use the tilde notation \(\widetilde{X}=\frac{X-\hat{\mu}_X}{\hat{\sigma}_X}\) for a random variable \(X\) with estimated mean \(\hat{\mu}_X\) and estimated standard deviation \(\hat{\sigma}_X\). Figure \ref{fig:correlation susceptibility regression} plots (on the original scale) the true correlation susceptibility values against their regression values.

\section{Conclusions}
\label{sec:conclusions}
This paper considers the behavior of the correlation of two quantities, the determinant \(\text{det}(A)\) and the eigenvalue stability \(\text{Re}(\lambda_\text{ls})\), associated with the system linearization \(A\) under the influence of perturbations of the system dynamics of the form equation \eqref{eq:pert signal}. These correlations are presented as indicators of the degree to which distinct signal inputs interact cooperatively versus acting independently. The necessity of normalization of the perturbation magnitudes leads to the introduction of the correlation susceptibility (equations \eqref{eq:uvar correlation detailed} - \eqref{eq:average uvar correlation susceptibility}).

Sections \ref{sec:sparse matrices} - \ref{sec:numerical arc analysis} suggest that increasing network arc density may decrease the determinant and eigenvalue correlation susceptibilities. Section \ref{sec:numerical analysis pert} confirms this suggestion via simulation for a class of quadratic dynamic systems, while also indicating similar trends with regard to network clustering and degree heterogeneity. Section \ref{sec:structural metrics} constructs a structural metric, the normalized arc density (equation \eqref{eq:normalized arc density}), which provides a potential unified explanation for the impacts of these distinct aspects of network structure.

The results with regard to (normalized) arc density are somewhat reminiscent of the dimensionality dependence of mean field theory in statistical physics \cite{goldenfeld1992lectures}. An Ising model spin is less sensitive to fluctuations of a nearby spin in a higher dimensional/higher connectivity lattice in a manner similar to the decreasing sensitivity of the response to one perturbation with respect to the influence of another in the dynamic systems studied here as normalized arc density increases.

Section \ref{sec:structural metrics} also introduces the eigenvalue contraction metric, which appears to offer a predictor of the correlation susceptibilities complementary to the normalized arc density. Both the network density and the eigenvalue contraction focus on the nodes most involved in the least stable eigenvector(s) of the linearization \(A\), which may be loosely thought of as the points at which the network is most susceptible to destabilization.

The results with regard to the normalized arc density and eigenvector contraction indicate that networks with lower arc density in the neighborhood of these focal nodes, and for which the fluctuations of the dynamic variables associated with these nodes are anticorrelated/out-of-phase with regard to the least stable dynamic modes, are more likely to respond to multiple signal inputs in a cooperative manner. It is intriguing to contemplate extension of this idea to develop more easily calculable (i.e., not linearization \(A\)-eigendecomposition-dependent) system structural metrics for potential use as predictors of the degree of interaction between distinct perturbations (such as, for example, an infection and a drug treatment).

\appendix
\appendixpage
\section{Generation of Directed Network Structures}
\label{sec:network generation}
Random digraph structures were generated by independently adding each arc \((i \rightarrow j)\) to the digraph with a fixed probability \(p_\text{arc} = \frac{d}{n-1}\), where the parameter \(d\) controls the average node in- and out-degree and \(n\) is the number of nodes in the network.

Directed scale-free networks with \(n\) nodes and \(2nd\) arcs were generated in a manner similar to the Barabasi preferential attachment mechanism \cite{barabasi1999emergence}. First, \((d+1)\) fully connected nodes were added. Then \(n-(d+1)\) nodes were added sequentially one-by-one, with \(d\) arcs added directed from the newly added node to old nodes and \(d\) arcs added from old nodes to the new node at each step. For each of these \(2d\) new arcs, the identity of the adjacent old node (whether it be tail or head of the new arc) was chosen at random with non-uniform probability proportional to the sum of the in- and out-degrees of the old node (subject to the constraint that no two arcs may connect the same pair of nodes with the same directionality).

Small-world digraphs with \(n\) nodes and \(2nd\) arcs (where \(d\) is even) were generated starting with a ring lattice in which each node is connected bi-directionally to the \(\frac{d}{2}\) nearest nodes in each direction, then rewiring each directed arc with probability \(p_\text{rw}\). Similar to the mechanism described by Watts, et. al. \cite{watts1998collective}, each rewired arc has one terminus (head or tail chosen randomly with uniform probability) changed to a node chosen randomly from all nodes in the system with uniform probability, again constrained to prevent any two arcs from sharing both the same tail and the same head.

\bibliographystyle{unsrt}
\bibliography{pap}

\begin{thebibliography}{1}

\bibitem{doroquez2006signal}
D.B. Doroquez and I.~Rebay.
\newblock {Signal integration during development: mechanisms of EGFR and Notch
  pathway function and cross-talk}.
\newblock {\em Critical reviews in biochemistry and molecular biology},
  41(6):339--385, 2006.

\bibitem{silva2008capturing}
J.~Silva and A.~Smith.
\newblock {Capturing pluripotency}.
\newblock {\em Cell}, 132(4):532--536, 2008.

\bibitem{wylie2009linked}
D.C. Wylie.
\newblock {Linked by loops: Network structure and switch integration in complex
  dynamical systems}.
\newblock {\em Physica A: Statistical Mechanics and its Applications},
  388(9):1946--1958, 2009.

\bibitem{schneider1989matrices}
H.~Schneider and G.P. Barker.
\newblock {\em {Matrices and linear algebra}}.
\newblock Dover Publications, 1989.

\bibitem{puccia1985qualitative}
C.J. Puccia and R.~Levins.
\newblock {\em {Qualitative modeling of complex systems}}.
\newblock Harvard Univ. Pr., 1985.

\bibitem{landin1989introduction}
J.~Landin.
\newblock {\em {An introduction to algebraic structures}}.
\newblock Dover Publications, 1989.

\bibitem{goldenfeld1992lectures}
N.~Goldenfeld.
\newblock {\em Lectures on phase transitions and the renormalization group}.
\newblock Addison-Wesley, Advanced Book Program, Reading, 1992.

\bibitem{barabasi1999emergence}
A.L. Barab{\'a}si and R.~Albert.
\newblock {Emergence of scaling in random networks}.
\newblock {\em Science}, 286(5439):509, 1999.

\bibitem{watts1998collective}
D.J. Watts and S.H. Strogatz.
\newblock {Collective dynamics of `small-world' networks.}
\newblock {\em Nature}, 393(6684):440--442, 1998.

\end{thebibliography}

\end{document}